\documentclass[
 reprint,
 aps,
 prl,
 superscriptaddress, 
 showpacs,
 longbibliography,
 amsmath,amssymb,
]{revtex4-2}

\usepackage[colorlinks=true, allcolors=blue]{hyperref}
\usepackage{gensymb}
\usepackage{float}
\usepackage{lipsum}
\usepackage{amsmath}
\usepackage{graphicx}
\usepackage{dcolumn}
\usepackage{bm}

\begin{document}

\title{Humble planar defects in SiGe nanopillars}

\author{Hongbin Yang}
\email{hongbin.yang@rutgers.edu}
\affiliation{Department of Chemistry and Chemical Biology, Rutgers University, Piscataway, New Jersey, 08854, USA}

\author{Shang Ren}
\author{Sobhit Singh}
\affiliation{Department of Physics and Astronomy, Rutgers University, Piscataway, New Jersey, 08854, USA}

\author{Emily M. Turner}
\author{Kevin S. Jones}
\affiliation{Department of Material Science and Engineering, University of Florida, Gainesville, Florida, 32611, USA}

\author{Philip E. Batson}
\affiliation{Department of Physics and Astronomy, Rutgers University, Piscataway, New Jersey, 08854, USA}

\author{David Vanderbilt}
\affiliation{Department of Physics and Astronomy, Rutgers University, Piscataway, New Jersey, 08854, USA}

\author{Eric Garfunkel}
\affiliation{Department of Chemistry and Chemical Biology, Rutgers University, Piscataway, New Jersey, 08854, USA}
\affiliation{Department of Physics and Astronomy, Rutgers University, Piscataway, New Jersey, 08854, USA}
\date{}

\date{\today}

\begin{abstract}
We report a new \{001\} planar defect found in SiGe nanopillars. The defect structure, determined by atomic resolution electron microscopy, matches the Humble defect model proposed for diamond. We also investigated several possible variants of the Humble structure using first principles calculations and found that the one lowest in energy was also in best agreement with the STEM images. The pillar composition has been analyzed with electron energy loss spectroscopy, which hints at how the defect is formed. Our results show that the structure and formation process of defects in nanostructured group IV semiconductors can be different from their bulk counterparts. 
\end{abstract}

\maketitle


Group IV semiconductors are ubiquitous in today’s electronics devices \cite{Paul2004SST}. Some of these materials also have promising characteristics for applications in quantum information processing \cite{Scappucci2020NRM, Leon2021Science}. Defects in these diamond cubic crystals may influence their properties in device applications, one key reason for the extensive study of their structure and formation process. One set of the most widely studied defects are the \{001\} planar defects in natural diamond \cite{Olivier2018NM}. Many structural models have been proposed in order to find a match with experiments. Amongst many other models \cite{Lang1964PPS,Miranda2004PRL}, Humble proposed a planar defect that can be regarded as resulting from the insertion of an entire layer of four-fold coordinated interstitial carbon atoms \cite{Humble1982}. Humble defects in Si have also attracted attention. Arai et al.~theoretically studied the self-interstitials in Si and found that forming defects with a Humble structure lowers the total energy significantly compared to free interstitials \cite{Arai1997PRL}. Goss et al.~further studied the bonding configuration of the Humble defect and proposed four other variations, and also calculated their energies, for both C and Si \cite{Goss2001PRB, Goss2002JOP, Goss2003PRB}.

\begin{figure*}[]
\includegraphics[width=17.2cm]{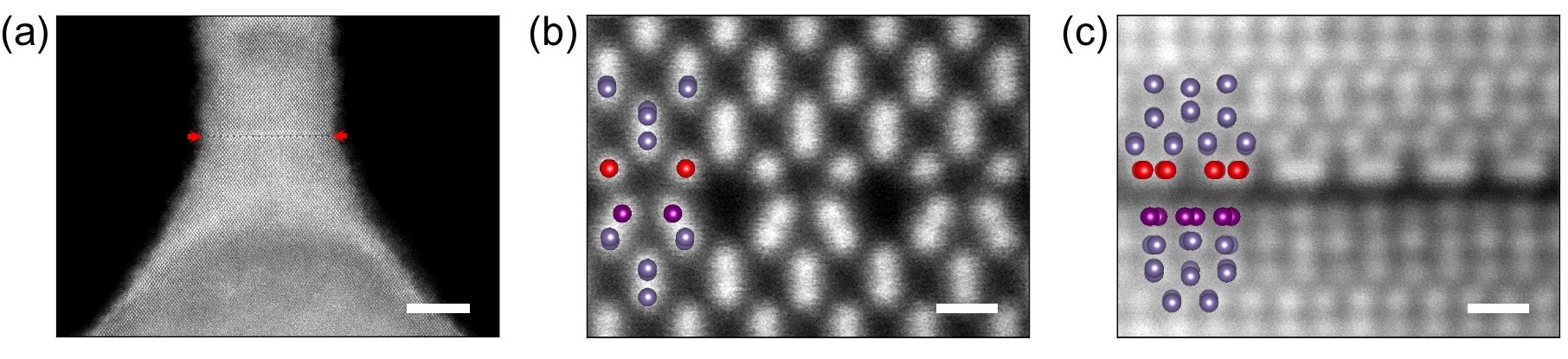}
\caption{Atomic structure of the SiGe Humble defect. (a) Low magnification HAADF-STEM images of the SiGe pillar. Scalebar: 10\,nm. The planar defect in this pillar is indicated by the two red arrows. (b) Atomic-resolution HAADF-STEM images of the defect viewed from [110] projection, and (c) from [210] projection, with the structural models shown superimposed. Scalebar: 3\,\AA.}
\label{fig:mesh1}
\end{figure*}

Experimentally, \{001\} planar defects were often observed when H was implanted into Si and Ge \cite{David2007JAP, Akatsu2005APL}. In ultrafast laser-annealed ion-implanted Si samples, defects in \{001\} planes appear as self-interstitial loops \cite{Marques2017PRL, Qiu2014NL}. Under thermal equilibrium, planar defects in Si and SiGe are primarily found along the \{113\} planes, and occasionally in \{111\} planes \cite{Fedina2000PRB}. Despite a large number of observations and calculations for defects in Si and Ge, the defect structure, preferred plane, and dimension are still not fully predictable. For the defects that have been observed earlier, only a few of them have been examined by atomic resolution high angle annular dark field (HAADF) imaging, which allows direct and quantitative comparison with computationally predicted structures \cite{Dudeck2013PRL}. In a recent electron microscopy study of planar defects in diamond, it was clearly shown that \{001\} platelet defects do not adopt the Humble structure \cite{Olivier2018NM}. Interestingly, the Humble structure can describe a \{001\} planar defect found in Ge \cite{Muto1995PML}. In Si, however, there is no clear evidence of Humble defects in any experiments. A number of calculations found that \{001\} and \{113\} oriented defects in Si have similar energies \cite{Goss2002JOP,Kapur2010PRB,Kapur2010PRB2}.

\begin{figure}[]
\includegraphics[width=8.6cm]{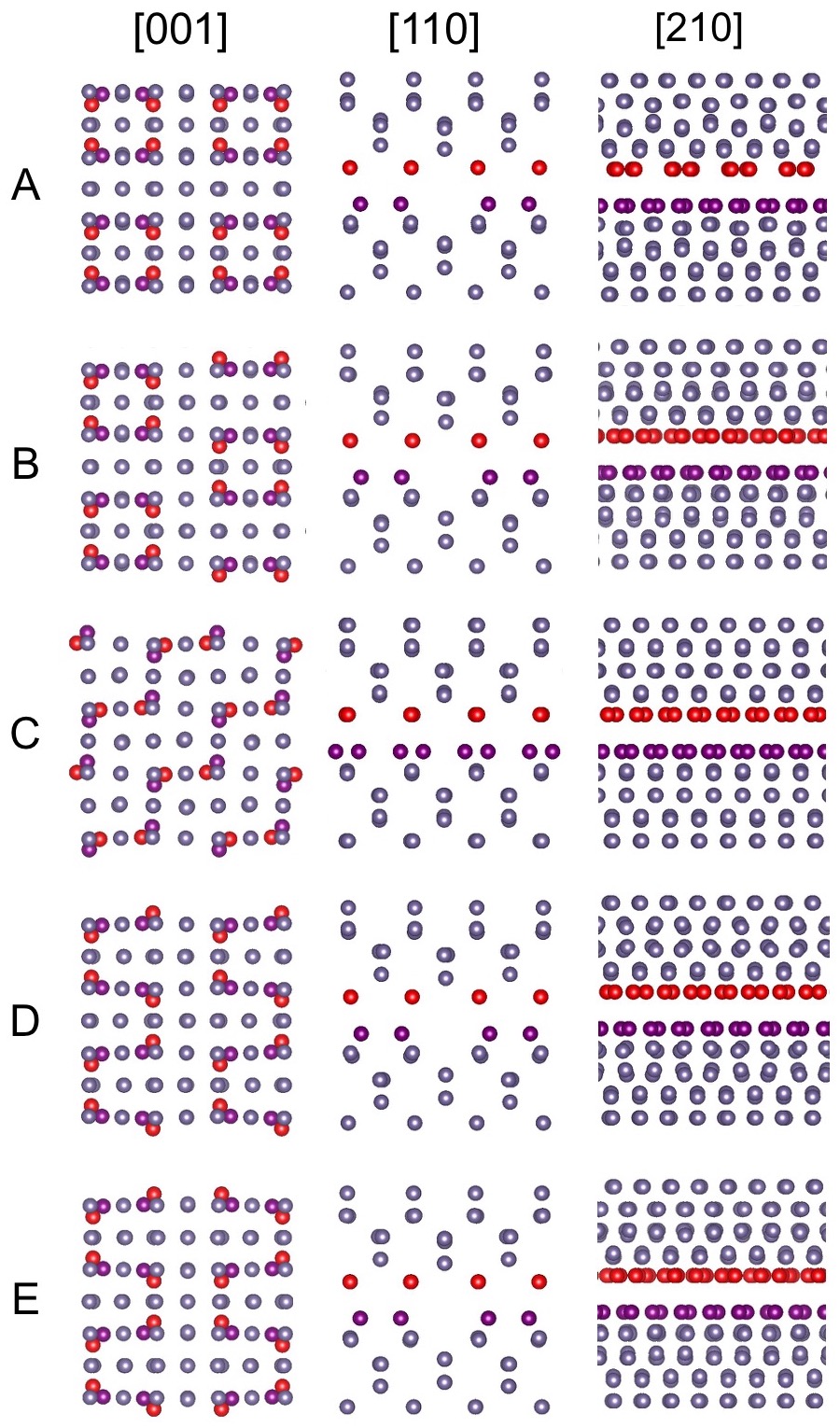}
\caption{Structural models of the five Humble defect proposed by Goss et al.~viewed from [001], [110], and [210] axis, defined relative to bulk lattice. Atoms shown in red and purple indicate the upper and lower atoms that are under significant bond angle distortion.}
\label{fig:mesh2}
\end{figure}

While bulk group IV materials and their defects have been extensively studied for decades, nanostructures of these materials are a subject of active research for their new and potentially useful electronic and optical properties~\cite{Zhu2009PRL,Bux2009AFM,Chan2010NanoLett,  Fadaly2020Nature}. During the growth or fabrication of these nanostructures, new structural phases and defects are occasionally encountered~\cite{Qiu2015SciRep, Dushaq2019SR, Kozhemiako2019Semiconductors, Bradac2019NatureCommunications, Fadaly2021NL}. The deliberate placement of defects in the nanostructures may offer yet another knob to tune properties or to create new classes of devices.

In this Letter, we report the first experimental observation of Humble defects in $\mathrm{Si_{0.2}Ge_{0.8}}$ (hereafter refered as SiGe). We have investigated the atomic structure of this defect by HAADF imaging in an aberration corrected electron microscope. The defect structure and atomic coordinates in experiments are in good agreement with density functional theory calculations. The local Si percentage and electronic structure across the defect have also been studied with high-resolution electron energy loss spectroscopy (EELS), which provides information on how these defects are formed.

The defects observed in this work are found in SiGe nanopillars formed by oxidation of Si/SiGe superlattices that are patterned into cylindrical rods \cite{Turner2021NL}. Oxidations were done under dry $\mathrm{O_2}$ at 900\degree C for 20\,minutes. Oxidation of SiGe results first in the formation of $\mathrm{SiO_2}$, as Ge oxides are much less stable than $\mathrm{SiO_2}$. When the desired oxidation time is reached, the wafer was quenched to room temperature within 1 minute. A TEM lamella was extracted by a focused ion beam lift-out process. We then remove the oxides by hydrofluoric acid etching \cite{Turner2020UM}. The remaining partially oxidized SiGe pillar was subsequently studied in this paper using a Nion UltraSTEM 100 aberration-corrected scanning transmission electron microscope (STEM).

One image of a SiGe pillar is shown in shown in Fig. \hyperref[fig:mesh2]{\ref{fig:mesh2}(a)}, where a planar defect forms parallel to the wafer surface; the defect appears uniform across this 17\,nm diameter pillar, and in many other similar pillars. Atomic-resolution imaging of this defect has been carried out in our STEM operating at 60\,kV, below the damage threshold \cite{Fedina2000PRB}. Atomic resolution HAADF images of the defect from two directions are shown in Fig. \hyperref[fig:mesh2]{\ref{fig:mesh2}(b)} and \hyperref[fig:mesh2]{\ref{fig:mesh2}(c)}. These images allow us to determine the defect structure and compare it with structural models proposed earlier for other group IV materials and structures.

For an extended two-dimensional Humble defect, the atomic arrangement at the defect core is not unique. Goss et al.~proposed several variations of the original Humble configurations. Figure \hyperref[fig:mesh1]{\ref{fig:mesh1}} shows all of these models as viewed from three major axes~\cite{Momma2011JAC}. The difference between these models primarily stems from the two layers of atoms shown in red and purple, which we will refer to as the defect core in the rest of this paper. Either the red or the purple layer can be considered as a layer of interstitial atoms, whose removal could result in a perfect bulk lattice. The atoms at the defect core form bonds within the a-b plane, but with nearest neighbors in different directions, as can be seen from the [001] view in Fig. \hyperref[fig:mesh2]{\ref{fig:mesh2}}. Within these models, all atoms in both bulk and defect core are 4-fold coordinated. The different bonding configuration at the defect core leads to large distortions in bond angles, different from the perfect tetragonal bonding geometry in an ideal bulk diamond cubic lattice, where all the bond angles are 109.5\degree. 

\begin{table}[]
\caption{\label{tab:example}DFT-calculated formation energy (eV/interstitial) for the five Humble models in pure Ge and $\mathrm{Si_{0.2}Ge_{0.8}}$.}
\begin{ruledtabular}
\begin{tabular}{llllll}
   Models & A & B &C &D &E\\
   \hline
   Ge & 0.309 & 0.354 & 0.422 & 0.363 & 0.360\\
   SiGe & 0.297 & 0.342 & 0.411 & 0.351 & 0.350\\
\end{tabular}
\end{ruledtabular}
\end{table}

\begin{figure}[]
\includegraphics[width=8.6cm]{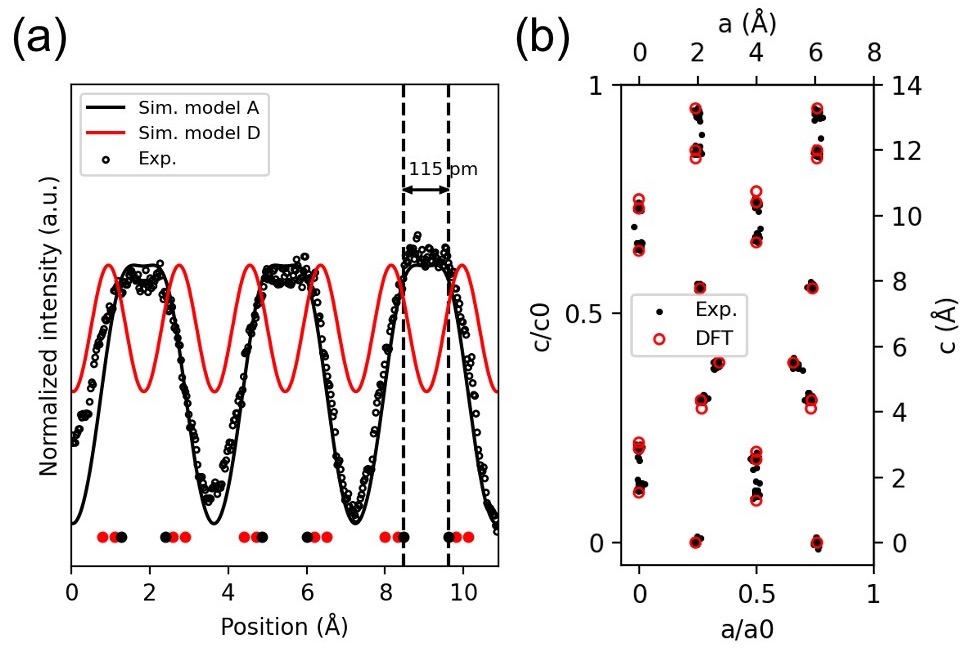}
\caption{Experimental defect structure compared with DFT-calculated ones. (a) HAADF intensities profiles of HAADF image in Fig.~1(b) for the upper layer of defect core compared with simulated ones from Humble models A and D. DFT-calculated atomic coordinates are shown as black and red dots for model A and D, respectively, at the bottom. (b) Experimental atomic coordinates from 2D gaussian fitting of HAADF images in Fig.~1 compared with DFT relaxed Humble model A.}
\label{fig:mesh3}
\end{figure}

To identify which of these variants of the Humble defect are present in our sample, we start with a preliminary screening of their energies using density-functional theory (DFT) calculations. For this purpose, formation energies of various Humble defect models are calculated using DFT. A 9-layer tetragonal unit cell is found to be sufficient in our calculations to capture the essential structural and energetic properties. In this work, all DFT calculations are performed using the Vienna Ab-initio Simulation Package (VASP) \cite{Kresse1996PRB} and the projector-augmented wave (PAW) \cite{Blochl1994PRB,Kresse1999PRB} method with Ge $\mathrm{4s^24p^2}$ and Si $\mathrm{4s^24p^2}$ pseudopotential valence configurations. The exchange-correlation used is the generalized-gradient-approximation (GGA) as parameterized by Perdew, Burke, and Ernzerhof (PBE) \cite{Perdew1996PRL}. The relaxation of the 9-layer Humble A structure is performed with a 6×6×4 Monkhorst-Pack \cite{Monkhorst1976PRB} k-mesh, and a corresponding k-mesh is used for the relaxation of Humble A to E structures. The in-plane lattice constants are fixed to the bulk value ($\mathrm{8\,\AA}$, $\mathrm{\sqrt{2}a_{SiGe}}$), and only the c lattice constant is relaxed. The convergence criteria for forces and energies are $\mathrm{10^{-3}\,eV⁄\r{A}}$ and $\mathrm{10^{-7}\,eV}$ respectively, and the cutoff energy for the plane-wave basis is 500 eV. The virtual crystal approximation (VCA) \cite{Bellaiche2000PRB} is used when simulating Humble defects in the $\mathrm{Si_{0.2}Ge_{0.8}}$ alloy. 

All structural models in Fig. \hyperref[fig:mesh2]{\ref{fig:mesh2}} show a relatively low formation energy, as shown in table I, with model A having the lowest energy in the case of both pure Ge and $\mathrm{Si_{0.2}Ge_{0.8}}$. The formation energy of \{001\} Humble defect in pure Si is 0.45\,eV per interstitial \cite{Goss2002JOP}, higher than those in pure Ge or Ge-rich SiGe (Table I), suggesting that forming \{001\} Humble defects in pure Si is energetically less favorable. However, formation energies of the \{001\} Humble defects in diamond is an order of magnitude higher than our results \cite{Goss2003PRB}. These results explain why Humble defects are only observed in SiGe or Ge but not in Si or C.

Next, having identified model A as a leading candidate based on theory, we turn to a direct determination of the structure based on STEM imaging. Despite having different atomic structure, model A, B, D, and E are almost identical when viewed from the [110] direction. Therefore, imaging only from [110] (the most commonly imaged axis for a diamond cubic crystal) is not sufficient to determine the defect atomic structure. As shown in Fig.~\hyperref[fig:mesh1]{\ref{fig:mesh1}(b)} and in Fig.~\hyperref[fig:mesh2]{\ref{fig:mesh2}}, we see five- and eight- member rings at the defect. The lower half of the defect resembles the symmetric dimers of a 2×1 reconstructed (001) surface of Si. To fully understand the atomic arrangements at the defect core, the defect has also been imaged along the [210] projection. This is done by tilting the specimen so that the pillar rotates along the c-axis 26.6\degree\, away from [110] projection. The corresponding image is shown in Fig.~\hyperref[fig:mesh1]{\ref{fig:mesh1}(c)}, where the upper layer of the defect core show dumbbells of atom columns. Comparing the HAADF intensities with simulated ones from DFT relaxed structure models in Fig.~\hyperref[fig:mesh0]{\ref{fig:mesh3}(a)}, we see that model A yields the best match with experiments. Within each dumbbell, the two atomic columns are separated by 115\,pm. And the distance between the dumbbells is 363\,pm. The subtle differences between model B, D, and E in their [210] view are too small to differentiate. Therefore, we only simulate HAADF intensity from model D to compare with experiment. We find that they show equally spaced atomic columns that are separated by 180\,pm, which does not match with what we observe in experiments. The lower layer of the defect core (purple atoms in Fig.~\hyperref[fig:mesh2]{\ref{fig:mesh2}}) in all models show atom columns that are separated by either 40 or 56pm, therefore are observed as single atom columns in Fig.~\hyperref[fig:mesh1]{\ref{fig:mesh1}(c)} and cannot be used to distinguish between model B, D, and E.

A more quantitative comparison between experiment and theory is shown in Fig.~\hyperref[fig:mesh3]{\ref{fig:mesh3}(b)}. From the HAADF images in Fig.~\hyperref[fig:mesh1]{\ref{fig:mesh1}}, we extract atomic positions by 2D gaussian fitting for each of the atomic columns. The fitting results are shown in Fig.~\hyperref[fig:mesh3]{\ref{fig:mesh3}(b)}, which overall match with DFT-relaxed atomic coordinates, although some subtle differences do exist, i.e., the bond distances in the purple layer are slightly larger in the experiment than calculated by DFT. We also see that even the atoms that are one layer away from the defect core show modulation in the c-axis direction. This shift is observed both in experiment and theory. 

\begin{figure}[]
\includegraphics[width=8.6cm]{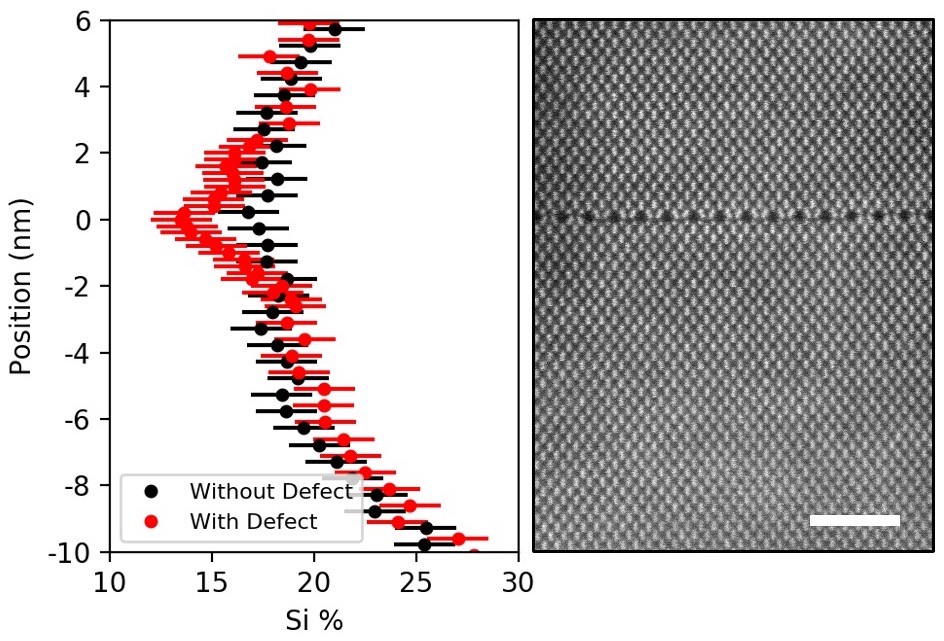}
\caption{Si percentage of SiGe pillars near the Humble defect. EELS line profiles of Si content across the defect in the area shown on the right. Same information for a SiGe pillar without the defect is shown in comparison. Scalebar: 2\,nm.}
\label{fig:mesh4}
\end{figure}

By combining atomic resolution ADF imaging with DFT calculations, we have completely determined defect structure. We conclude that the symmetry of the original Humble defect (model A) applies to the one reported here in SiGe, and that atomic coordinates of the DFT relaxed structure match well with the experimental results. This defect is equivalent in its [110] and [1-10] views, both of which are observed in the HAADF image in Fig.~\hyperref[fig:mesh4]{\ref{fig:mesh4}}, where the upper and lower layers of the defect core alternate in the a-b plane. In addition to this predominant defect that we observe in many pillars, we also note that \{001\} defects with a different structure than model A, and even a \{113\} defect have also been observed, but much less frequently (only in one pillar).

As mentioned earlier, the defects are observed after oxidation of Si/SiGe nanopillars. This process is unusual and entirely different from earlier work on planar defects in Si and Ge, in which the defect creation process usually involved ion implantation or high energy electron irradiation \cite{Muto1995PML, Fedina2000PRB}. Creation and diffusion of point defects, such as self-interstitials and vacancies, are often responsible for the formation of extended defects. 

Here, we propose that the point defects are created in the form of vacancies, as a result of Si out-diffusion to form $\mathrm{SiO_2}$. During high temperature oxidation, Si atoms are preferentially oxidized \cite{LeGoues1989JAP} while Ge atoms diffuse into the $\mathrm{Si_{0.3}Ge_{0.7}}$ film \cite{Brewer2017NL}. Therefore, the Si percentage in the SiGe layer drops by about 10\% to 20\% after oxidation, and the initially sharp Si/SiGe interface become one with graded Si percentage, as quantified by EELS in Fig.~\hyperref[fig:mesh4]{\ref{fig:mesh4}}. Very close to the defect, we see a further decrease in Si content to lower than 20\%. This suggests that the defect occurs with predominantly Ge-Ge bonding. This tendency of having less Si at the defect core is also supported by DFT calculations. When substituting one Si atom into the 13-layer Humble unit cell, the defect core is the least preferred location. With the fast cooling after oxidation, point defects aggregate and eventually form the defects we observe here. It is also interesting to note that, although there are several alternating layers of SiGe/Si on the Si substrate, the Humble defects were only found in the first SiGe layer after oxidation. This suggests that strain from the Si substrate maybe also be involved in forming such a planar defect. 

The electronic structure of this defect has been investigated by high resolution electron energy loss spectroscopy and density functional theory, and will be reported elsewhere \cite{Ren2021}.

In summary, we found that \{001\} Humble planar defects can form in SiGe nanopillars after dry oxidation and quenching, a process that was not previously known to create such defects. By combining atomic resolution electron microscopy and first-principles calculations, the defect structure, composition and energetics have been studied in detail, and compared with previous studies on other group IV semiconductors. Future study of the compositional, pillar size, and processing condition dependence of the Humble defect will provide further insight into the formation mechanism.

\section{Acknowledgement}
S. R. and D. V. are supported by NSF Grant DMR-1954856.

\bibliography{Humble_defect}
\bibliographystyle{apsrev4-2}

\end{document}